\documentclass[pdflatex,sn-mathphys-num]{sn-jnl}

\usepackage{graphicx}%
\usepackage{multirow}%
\usepackage{amsmath,amssymb,amsfonts}%
\usepackage{amsthm}%
\usepackage{mathrsfs}%
\usepackage[title]{appendix}%
\usepackage{xcolor}%
\usepackage{textcomp}%
\usepackage{manyfoot}%
\usepackage{booktabs}%
\usepackage{algorithm}%
\usepackage{algorithmicx}%
\usepackage{algpseudocode}%
\usepackage{listings}%

\usepackage{fontawesome5}
\usepackage{etoolbox}

\theoremstyle{thmstyleone}%
\theoremstyle{thmstyleone}%

\theoremstyle{thmstyletwo}%

\theoremstyle{thmstylethree}%
\makeatletter
\def\blfootnotetext{\xdef\@thefnmark{}\@footnotetext}
\makeatother

\begin{document}

\title[ ]{Algorithmic exploration of the unit distance problem in the rational plane}

\author*{\fnm{Panteleimon} \sur{Rodis}
\href{https://orcid.org/0000-0001-9169-8202} 
\orcid{}}

\abstract{
This paper presents reproducible experimental evidence on unit-distance graph density. Our approach is based on a novel algorithmic exploration of the rational plane for the generation of unit-distance graphs. An efficient algorithm for this utility must perform a local-breadth search on a bounded and finite set of elements and generate a graph that potentially encompasses the general properties of a unit-distance graph, not affected by restrictions on its generation. To this end, we show that our approach accomplishes this purpose by overcoming the limitations of grid-based structures, used in the literature for generating unit-distance graphs. When applied to large unit-distance graphs, our method generates large graphs with a scaling exponent of 1.17, that surpasses recent theoretical predictions on graph density growth.
}


\maketitle

\blfootnotetext{

\small 
\faEnvelope~rodis@uom.edu.gr,~~
\href{https://pantelisrodis.blogspot.com/}{\faLink}
\hfill September 2026

}

\section{Introduction}

The unit distance problem, introduced by Paul Erdős \cite{erdos1946sets}, concerns the maximum possible number of unit-distance pairs in a set of $n$ points in the plane. Erdős estimated a lower bound of $n^{1+ c/ \log \log n}$ for the maximum number of unit-distance pairs.

Recent results developed by an OpenAI model \cite{openai} and further improved by Will Sawin \cite{sawin2026explicit} have raised the lower bound barrier to $n^{1.014}$, while the upper bound was established at $O(n^{4/3})$ by Spencer, Szemerédi and Trotter \cite{spencer1984unit}. It is a field of active study including a mixture of theoretical and experimental techniques \cite{alon2026remarks}, \cite{emmerich2026optimizing}. The most common approach to the problem, also utilized in the aforementioned studies, is the definition of a geometrical structure and the study of its iterative development to evaluate how infinite instances of this structure can be combined to form high dimensional, dense unit-distance graphs. The results depend on the chosen structure and the method of combining infinite instances, raising doubts about their generalization.

Experimental investigations of the problem are confined by its computational hardness. Working in the plane of reals raises imprecision issues, as computing distances that involve irrational numbers can only provide approximate results. In cases where researchers use integer representations of the plane, e.g. the Moser lattice \cite{engel2025diverse}, they face inherent restrictions of the representation itself which affect generality.

To overcome these obstacles, the presented method develops a greedy exploration of the plane to discover unit distant points without some predefined search fashion. To overcome precision issues, we limit the search to the plane of rational numbers, where $\forall q \in \mathbb{Q}, \exists a, b \in \mathbb{Z} \rightarrow q=a/b \land b \neq 0$ and distance calculation can be reduced to integer multiplications, additions and subtractions. It is trivial to show that any graph defined in $\mathbb{Q}^2$ has infinitely many isometric embeddings in $\mathbb{R}^2$. As such, the results of our approach extend beyond the rational plane.

Still, rationals are dense in the plane and as such to overcome computational hardness, the search is confined to a progressively expanding space and a limited set of denominators for computing rational coordinates. These constraints enabled the development of an algorithm capable of efficiently generating highly dense unit-distance graphs, providing a clear insight into the problem.

The main contribution of this work is a novel method of constructing large unit-distance graphs, of more than 2.4 million nodes in our implementation, that achieve a scaling density that exceeds $n^{1.17}$ and surpasses theoretical lower bound estimation. We also provide empirical evidence demonstrating that grid-based graph generation methods tend to conform to theoretical limitations of lower bounds. The source code of the algorithmic implementation and the experimental results are open-sourced at \cite{repository}, encouraging reproducibility and further exploration by other researchers.

In the next section, we detail the calculation methods and constraints used to ensure the computational efficiency of our algorithm, we provide a finite graph formulation of the problem suitable for our experimental approach and provide a description of the search algorithm. In section~\ref{results}, we present the experimental results of the greedy algorithm. Subsequently, as a means of comparison, we present a heuristic variation of our method that develops a grid-based graph. Section~\ref{conclusions} summarizes our conclusions and discusses future directions of this research.

\section{Algorithmic exploration}
\label{efficiancy}

Working on the rational plane enables the efficient computation of Euclidean distances, as shown in the next subsection. Nevertheless, this assumption also has implications for graphs on the plane of real numbers. It is well established that any graph $G$ with rational coordinates has an uncountably infinite number of isometric embeddings $G'$ in the real plane under the following transformations:
\begin{itemize}
    \item For $a,b \in \mathbb{R}$, the coordinates $(x,y)$ of every node of $G$ are transformed to $(x+a,y+b)$ and generate isometric graph $G'$ in $\mathbb{R}^2$. The possible values for $a,b$ are uncountably infinite so are the possible isometric embeddings $G'$.
    \item Graph $G$ can be rotated by a real angle $\theta \in (0, 2\pi)$ and generate isometric embedding $G'$. The values of $\theta$ are also uncountably infinite.
\end{itemize}

Note that the isometric graphs are still governed by the intrinsic properties of graphs defined in $\mathbb{Q}^2$.

Algorithmic exploration of the unit distance problem yields valuable insights into the complexity of its structure and potential. Nevertheless, algorithmic approaches inevitably explore finite structures. As such, we can safely prove results for a finite version of the problem termed as Finite Unit-Distance Graph (FUDG) and formulated as follows:
\begin{equation*}
    \text{FUDG}(n,k) = \{G(V,E)~\vert~G\text{ is a unit-distance graph, where }\vert V \vert>n \text{ and } \vert E \vert>n^k\}
\end{equation*}

In section~\ref{results} we prove the existence of FUDG$(2.4\cdot 10^6,1.17)$.

\newpage

\subsection{Euclidean distance calculation in the rational plane}
\label{distancecalculation}

Working with rational numbers circumvents computational precision issues under the following considerations.

For the euclidean distance of any two points $p_1 (x_1, y_1)$ and $p_2(x_2, y_2)$ in the plane, we seek to validate if  $\sqrt{(x_1 - x_2)^2 + (y_1 - y_2)^2} = U$, for constant $U$. 
Squaring both sides simplifies the equation to $(x_1 - x_2)^2 + (y_1 - y_2)^2 = U^2$.

Given that every rational number is equal to a fraction of integers:

\begin{equation*}
x_1 = \frac{s_1}{s_2}, y_1 = \frac{n_1}{n_2}, 
x_2 = \frac{u_1}{u_2}, y_2 = \frac{v_1}{v_2}
\text{, where~} s_i, u_i, n_i,v_i \in \mathbb{Z}
\end{equation*} 

the distance equation is formed as follows: 

\begin{equation*}
\left(\frac{s_1}{s_2} - \frac{u_1}{u_2}\right)^2 + \left(\frac{n_1}{n_2} - \frac{v_1}{v_2}\right)^2 = U^2 \Leftrightarrow
\end{equation*}

\begin{equation}
\label{eq:1}
\left(\frac{s_1 \cdot u_2 - u_1 \cdot s_2}{s_2 \cdot u_2}\right)^2 +
\left(\frac{n_1 \cdot v_2 - v_1 \cdot n_2}{n_2 \cdot v_2})\right)^2 = U^2
\end{equation} \\

we set:

$x_{num} = s_1 \cdot u_2 - u_1 \cdot s_2$

$x_{den} = s_2 \cdot u_2$

$y_{num} = n_1 \cdot v_2 - v_1 \cdot n_2$

$y_{den} = n_2 \cdot v_2$ \\

and substitute to equation~\ref{eq:1}:

\[\left(\frac{x_{num}}{x_{den}}\right)^2+ \left(\frac{y_{num}}{y_{den}}\right)^2 =\frac{x_{num}^2}{x_{den}^2} + \frac{y_{num}^2}{y_{den}^2} = U^2 \Leftrightarrow \]

\[\left(\frac{x_{num}^2}{x_{den}^2} + \frac{y_{num}^2}{y_{den}^2}\right) \cdot (x_{den}^2 \cdot y_{den}^2) = U^2 \cdot (x_{den}^2 \cdot y_{den}^2) \Leftrightarrow \]

\[\left(\frac{x_{num}^2}{{x_{den}^2}} \cdot {x_{den}^2} \cdot y_{den}^2\right) + \left(\frac{y_{num}^2}{{y_{den}^2}} \cdot x_{den}^2 \cdot {y_{den}^2}\right) = U^2 \cdot x_{den}^2 \cdot y_{den}^2\] 

the final formulation is:

\[x_{num}^2 \cdot y_{den}^2 + y_{num}^2 \cdot {x_{den}^2} = U^2 \cdot x_{den}^2 \cdot y_{den}^2\]

We set $U=1$ for unit-distance pairs of points. The calculation of euclidean distance over the rational plane is then reduced to integer multiplications, additions and subtractions. In this design, we circumvent the precision issues of computer hardware associated with floating point representations.

\subsection{Graph expansion algorithm}
\label{algorithm}

\begin{algorithm}
\caption{Greedy Fast Search}
\label{alg:search}
\begin{algorithmic}[1]
 \renewcommand{\algorithmicrequire}{\textbf{Input:}}
 \renewcommand{\algorithmicensure}{\textbf{Output:}}
\Require seed nodes, denominators set $D$, target $T$, parameter $h$, empty graph $G(V,E)$, number of iterations
\Ensure graph $G$

\item[]
\State run \textbf{main}()
\item[]
\State \textbf{Procedure main}()
\State add seed nodes to $V$
\For {each iteration}
\State $i = 0$
\While {i $< \vert D \vert$}
    \State run \textbf{pass}($D(i)$)
    \State $i++$
\EndWhile
\EndFor

\item[]
\State \textbf{Procedure pass}($d$)
\State calculate current circular boundary $r+h$
\State compute set $H$ of numbers $g_x/d$ within circle coordinates
\State empty set $P$
\For {$x =0$ to $\vert H \vert$}
    \For {$y =0$ to $\vert H \vert$}
        \If {point $(H[x], H[y])$ is within circle}
        \If {point does not coincide with existing points}
            \State add point to $P$
        \EndIf
        \EndIf
\EndFor
\EndFor

\For {every $p$ in $P$}
 \For {every $v$ in $V$}
    \If {euclidean distance $(p, v) = 1$}
        \State add $p$ to $V$
    \EndIf
 \EndFor
\EndFor

\end{algorithmic}
\end{algorithm}

The search is initiated at one or two seed nodes and the rest of the procedures are executed iteratively. 

We define a circular search space to achieve uniform distribution across all dimensions. In every step, the circle radius $r$ that is defined by the node of longer euclidean distance from the center of the circle, is further increased by $h$ to achieve a progressive and efficient expansion of the search space. 

Parameter $h$ must be chosen so that the graph expansion does not surpass space growth, thereby avoiding the termination of the procedure. Concurrently, graph expansion must be the minimum possible to avoid redundant computations, optimize search efficiency and develop a highly dense graph by maximizing node proximity.

Even in a bounded search space, rational numbers are still dense. To establish a finite computational procedure, we define a finite set of denominators $D$. The search procedure, for every $d \in D$ evaluates all the values of the form $a = x/d$ and $b = y/d$, where point $(a,b)$ is an interior point of the circle. Every newly discovered point that forms a unit-distance pair with an existing point in the graph, is added in the set of nodes of the graph.

Graph growth is not based on a pattern or the repetition of a predefined structure but in a greedy search of the plane. Nevertheless, the limitation of expanding the graph only in the rational plane, promotes the development of symmetrical structures that exhibit high connectivity in large scale, as we demonstrate in section~\ref{results}.

Algorithm~\ref{alg:search} outlines the graph expansion procedure in pseudocode. The computational demands of the procedure mandate the use of advanced hardware, specifically GPUs. The deterministic nature of the algorithm enables it to be executed in discrete parts on demand. To this end, the code was further optimized using AI code assistants. The final version includes a memory-efficient filter designed to prune the vast majority of non unit-distance candidate pairs and a unit-distance verification method via modular arithmetic. The algorithm builds the set of nodes of the graph, where each node participates at least in one unit-distance pair. The set of edges can be computed independently for any desired subset of the graph.

\section{Experimental results}
\label{results}

The greedy graph expansion algorithm was evaluated through a large-scale experimental application. We also developed a heuristic variation of the algorithm to gain more insight into the procedure. The heuristic applies more constraints to the search in the rational plane so that the added points eventually develop a structure and the graph expands by combining multiple copies of this structure.

The results presented in this section demonstrate the advantages of the greedy search within the plane, which surpasses theoretical limitations in comparison to the restrictive heuristic method that tends to conform to theoretical lower bounds. The source code of both algorithmic implementations and the experimental results are available at \cite{repository}.

\subsection{Greedy search on the plane}
\label{greedy}

We set the following parameters for algorithm~\ref{alg:search}:

\begin{itemize}
    \item Parameter $h=0.01$.

    \item Seed nodes $s_1(0/ 1, 0/1)$ and $s_2(-1/1, 0/1)$.

    \item Set of denominator $D = \{5, 25, 50, 65, 85, 100, 125, 130, 145, 170, 185, 200\}$, each composed only of 2 and $4k+1$ primes.
\end{itemize}

\begin{figure}[H]
  \centering
  \begin{minipage}{0.5\textwidth}
    \centering
    \includegraphics[width=\textwidth]{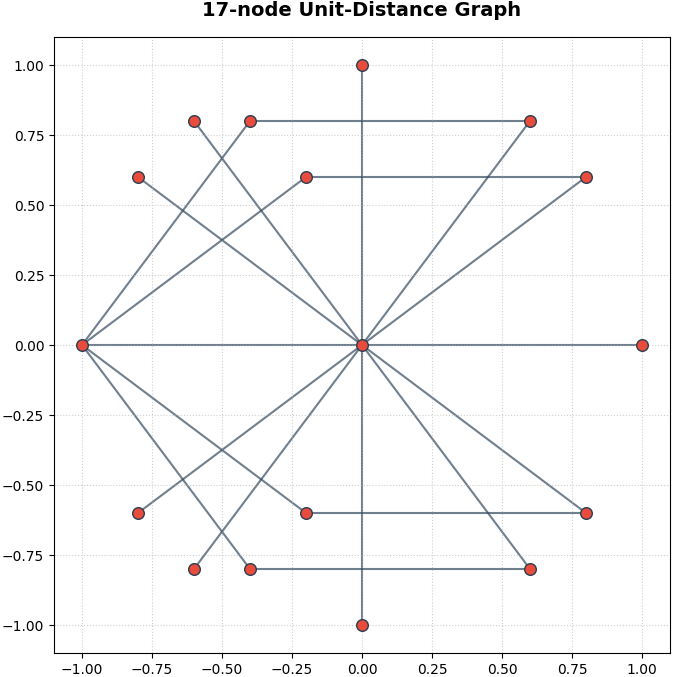}
    \caption{First pass.}
    \label{fig:image1}
  \end{minipage}%
\hfill
  \begin{minipage}{0.5\textwidth}
    \centering
    \includegraphics[width=\textwidth]{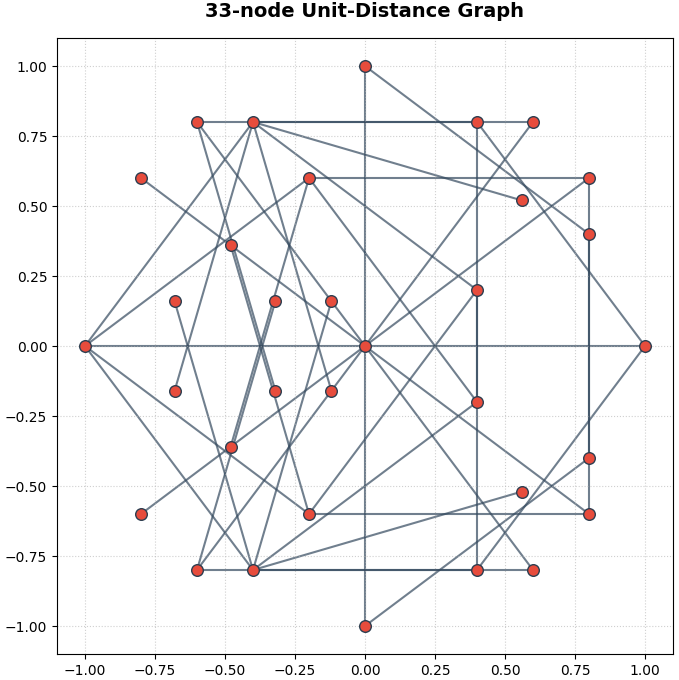}
    \caption{Graph Expansion to 33 Nodes.}
    \label{fig:image2}
  \end{minipage}
\end{figure}

The algorithm after 20 iterations, which required in total more than 30 hours to run, generated a highly dense graph of 2,411,001 nodes and 33,237,700 edges that achieves a scaling exponent that exceeds 1.1785. 
Whether this scaling behavior holds as the graph continues to grow is a subject of further experimental or theoretical investigation. 
Figures~\ref{fig:image1},~\ref{fig:image2} illustrate the initial stages of graph development, where nodes form symmetries across multiple directional axes.

\begin{figure}[H]
  \centering
  \begin{minipage}{0.5\textwidth}
    \centering
    \includegraphics[width=\textwidth]{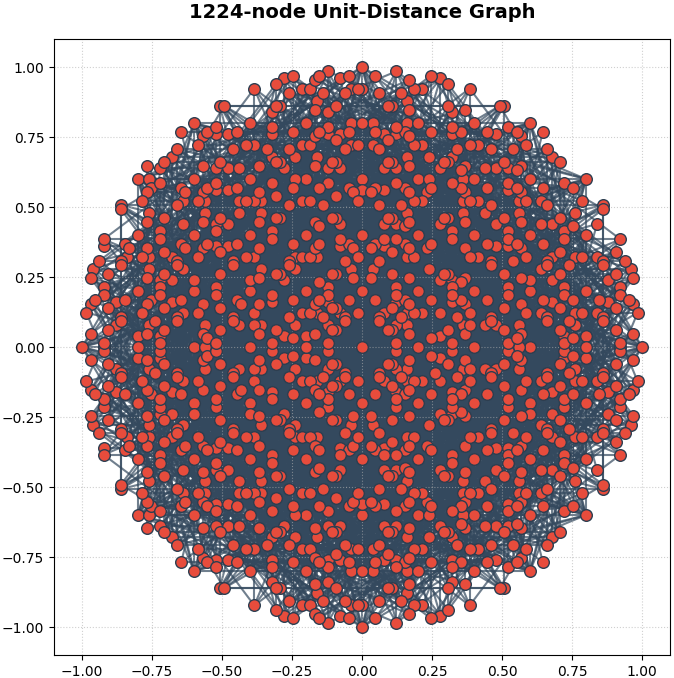}
    \caption{Highly symmetrical dense graph.}
    \label{fig:image3}
  \end{minipage}%
  \hfill
  \begin{minipage}{0.5\textwidth}
    \centering
    \includegraphics[width=\textwidth]{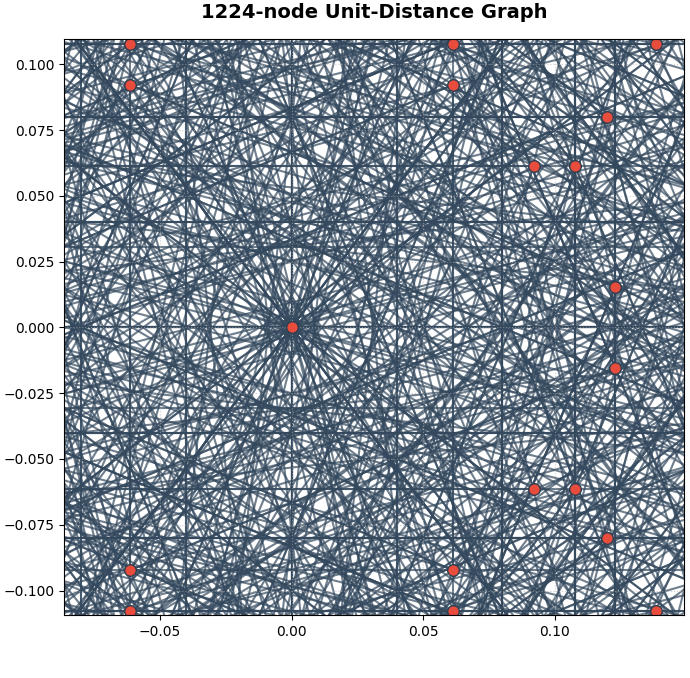}
    \caption{A highly complex network.}
    \label{fig:image4}
  \end{minipage}
\end{figure}

Figures~\ref{fig:image3}~and~\ref{fig:image4} illustrate the development of the graph up to 1,224 nodes; a larger instance of the graph becomes indistinguishable when plotted. The graph, at this stage, is confined to a restricted area of approximately 4 square units, resulting in a remarkably dense cluster. Particularly Figure~\ref{fig:image4}, displays complex symmetrical structures that become difficult to separate visually.

\begin{figure}[H]
  \centering
  \begin{minipage}{0.5\textwidth}
    \centering
    \includegraphics[width=\textwidth]{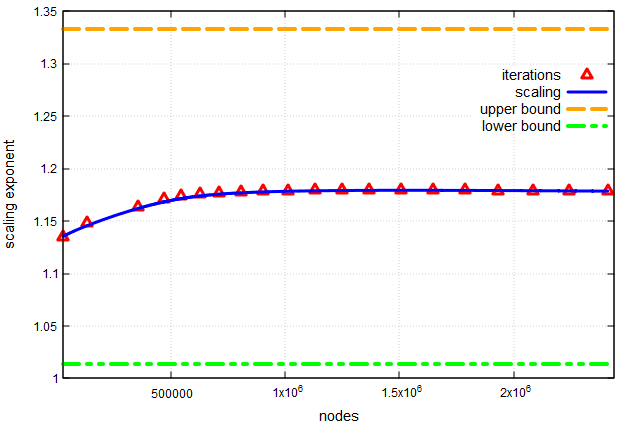}
    \caption{Greedy algorithm scaling exponent.}
    \label{fig:asymptotic}
  \end{minipage}%
  \hfill
  \begin{minipage}{0.5\textwidth}
    \centering
    \includegraphics[width=\textwidth]{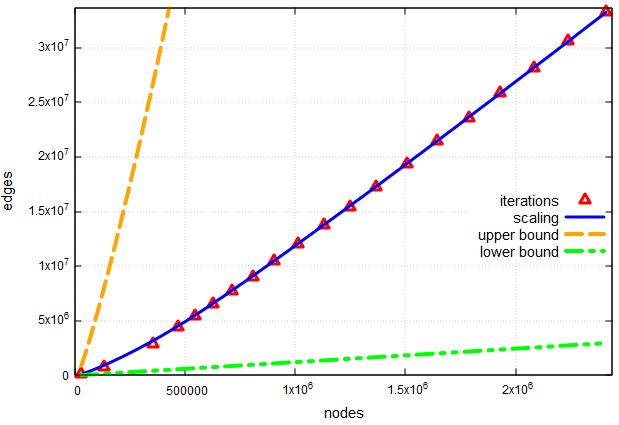}
    \caption{Edge growth trajectory.}
    \label{fig:edges}
  \end{minipage}
\end{figure}

\begin{figure}[H]
  \centering
  \begin{minipage}{0.5\textwidth}
    \centering
    \includegraphics[width=\textwidth]{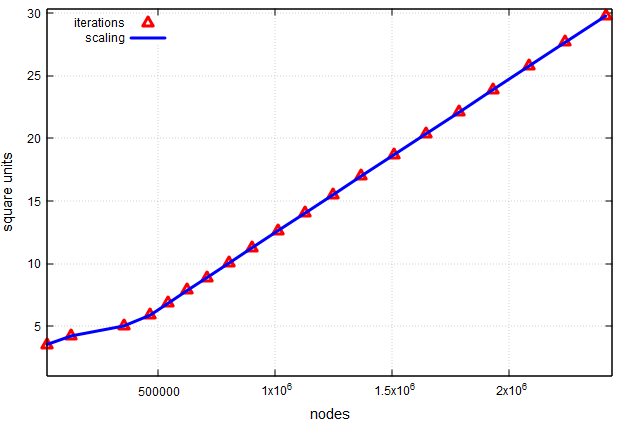}
    \caption{Spatial extent.}
    \label{fig:reas}
  \end{minipage}%
  \hfill
  \begin{minipage}{0.5\textwidth}
    \centering
    \includegraphics[width=\textwidth]{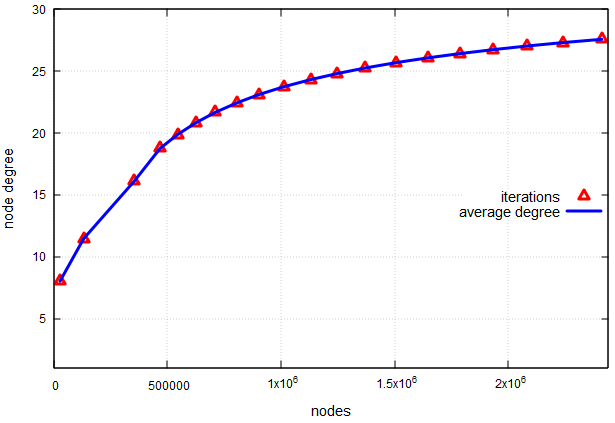}
    \caption{Average node degree.}
    \label{fig:degrees}
  \end{minipage}
\end{figure}

The scaling exponent of the graph, illustrated in Figure~\ref{fig:asymptotic}, reaches a plateau at 1.17, although it fluctuates slightly around 1.178. The development of the edges follows a superpolynomial rate in Figure~\ref{fig:edges}. The bounded search area maintains a restricted size, indicating the efficiency of the algorithm, Figure~\ref{fig:reas}. Figure~\ref{fig:degrees} illustrates the average node degree. As expected, node degrees increase as newly added nodes form multiple unit-distance node pairs.

\subsection{Heuristic grid-based graph expansion}

The parameters for the heuristic search, outlined in algorithm~\ref{alg:search-dia}, are set as follows

\begin{itemize}
    \item Parameter $h=\sqrt{2}$.

    \item Seed node $s_1(0/ 1, 0/1)$.

    \item Set of denominator $D = \{1, 2, 3, 4, 5\}$.
\end{itemize}

The algorithm varies from the greedy approach. It utilizes a heuristic to find the single most suitable node to add to the graph in every iteration. To this end, the set of candidate points is evaluated to find the point that achieves the highest connectivity with the existing nodes in the graph.

A tiebreaker mechanism is introduced to resolve cases where multiple points have the maximum number of connections. The mechanism selects the point with the highest sum of denominators of its rational coordinates.

These variations are reflected in the GPU optimization. The algorithm is less demanding in terms of memory consumption, although it is slower in terms of graph expansion. As such, the optimized implementation of the heuristic algorithm utilizes a distinct approach to maximize GPU efficiency.

\begin{algorithm}
\caption{Heuristic Search}
\label{alg:search-dia}
\begin{algorithmic}[1]
 \renewcommand{\algorithmicrequire}{\textbf{Input:}}
 \renewcommand{\algorithmicensure}{\textbf{Output:}}
\Require seed node, denominators set $D$, parameter $h$, empty graph $G(V,E)$, number of iterations
\Ensure graph $G$

\item[]
\State add seed node to $V$
\For {each iteration}
    \State calculate current circular boundary $r+h$
    
    \For{every $d$ in $D$}
        \State compute valid point coordinates within circle
    \EndFor
    \State compute candidate points and number of connections to nodes in $V$
    \State add candidates in set $P$
    \State current-max-connections$=0$
    \State current-best = none
    \For{every $i$ in $P$}
        \If{connections of $i >$ current-max-connections}
            \State update current-max-connections
            \State current-best = $i$
        \ElsIf{connections of $i$ = current-max-connections}
        \If {denominators sum of $i >$ current-best}
            \State current-best = $i$
        \EndIf
        \EndIf
    \EndFor
    \State add current-best to $V$
\EndFor

\end{algorithmic}
\end{algorithm}

\begin{figure}[H]
  \centering
  \begin{minipage}{0.5\textwidth}
    \centering
    \includegraphics[width=\textwidth]{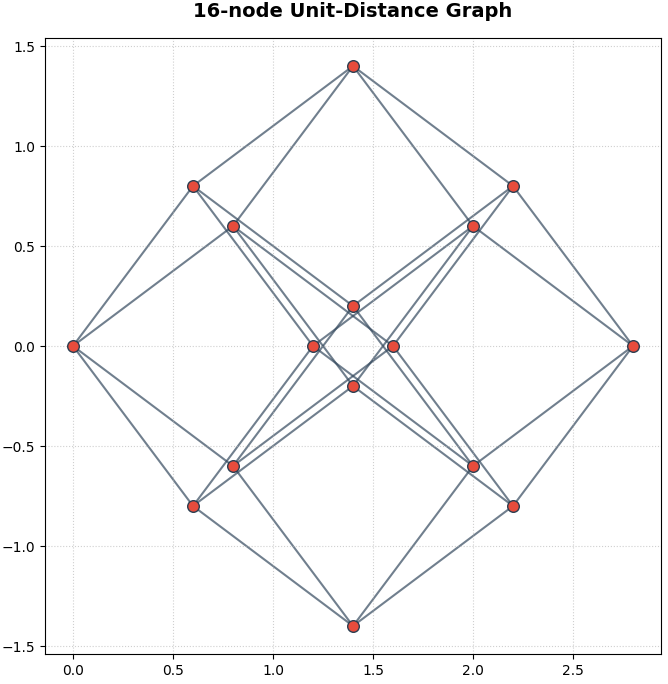}
    \caption{Initial graph formation resembling a diamond structure.}
    \label{fig:image7}
  \end{minipage}%
  \hfill
  \begin{minipage}{0.5\textwidth}
    \centering
    \includegraphics[width=\textwidth]{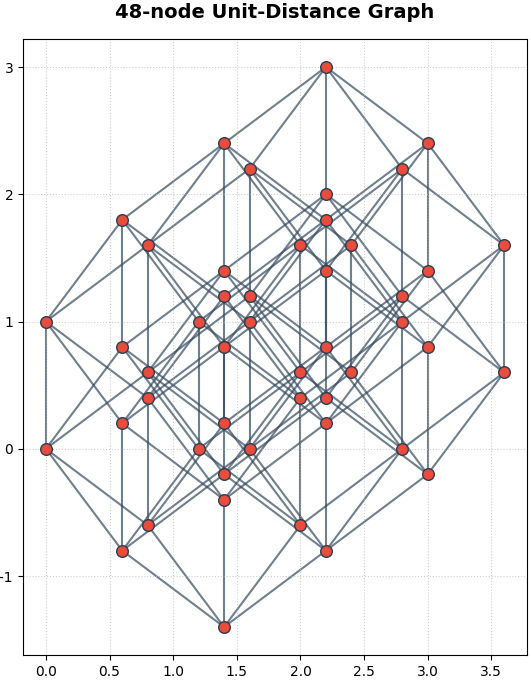}
    \caption{The graph expands by interconnecting multiple copies of the same pattern.}
    \label{fig:image8}
  \end{minipage}
\end{figure}

Figures~\ref{fig:image7}~and~\ref{fig:image8} illustrate the progressive development of the graph under these configurations. Initially, the algorithm generates a diamond-like pattern and subsequently interconnects copies of this pattern.

\begin{figure}[H]
  \centering
  \begin{minipage}{0.5\textwidth}
    \centering
    \includegraphics[width=\textwidth]{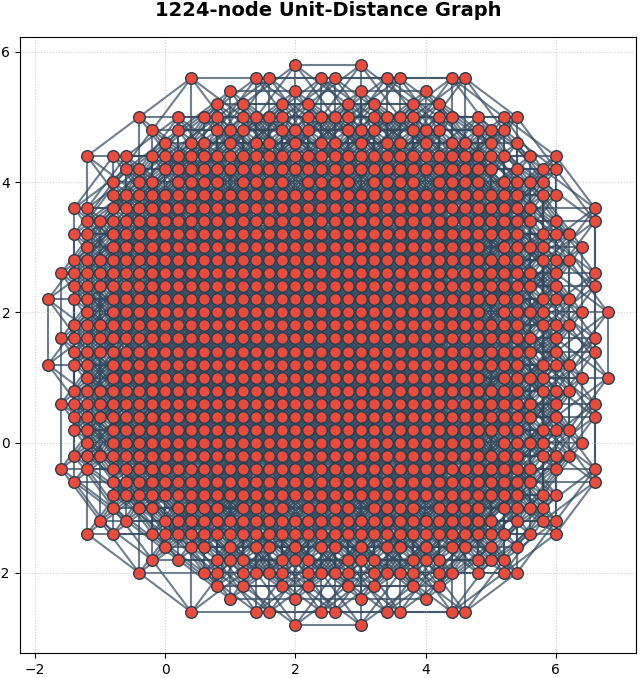}
    \caption{The algorithm generates a grid.}
    \label{fig:image9}
  \end{minipage}%
  \hfill
  \begin{minipage}{0.5\textwidth}
    \centering
    \includegraphics[width=\textwidth]{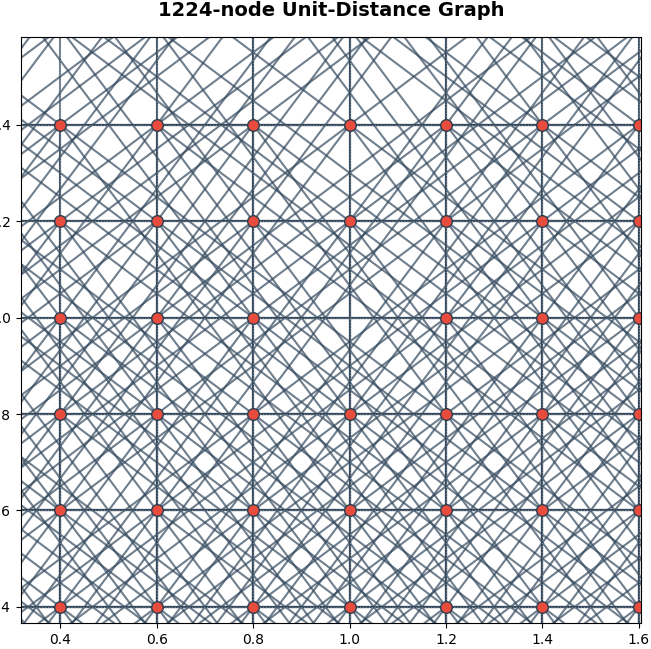}
    \caption{The grid forms simple patterns.}
    \label{fig:image10}
  \end{minipage}
\end{figure}

As the procedure progresses, the algorithm essentially builds a grid of the repeated pattern, as illustrated in Figure~\ref{fig:image9}, which resembles the techniques used in the theoretical approach to the problem. The expanded 1224-node graph consists of a symmetrical network (Figure~\ref{fig:image10}) that is significantly less complex than the greedy generated graph. The heuristic graph is clearly less dense, distributed across a broader space.

\begin{figure}[H]
  \centering
  \begin{minipage}{0.5\textwidth}
    \centering
    \includegraphics[width=\textwidth]{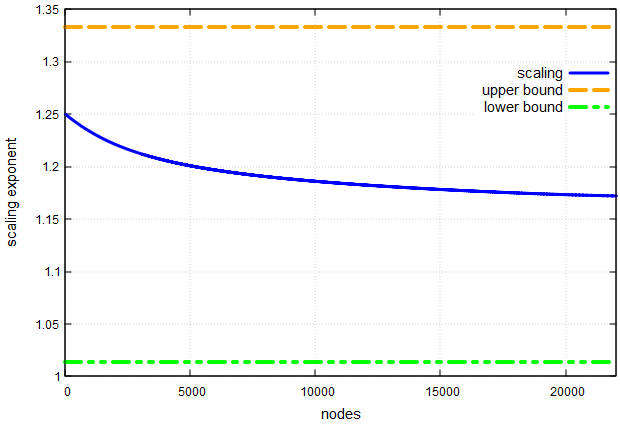}
    \caption{Scaling exponent.}
    \label{fig:asymptotic-dia}
  \end{minipage}%
  \hfill
  \begin{minipage}{0.5\textwidth}
    \centering
    \includegraphics[width=\textwidth]{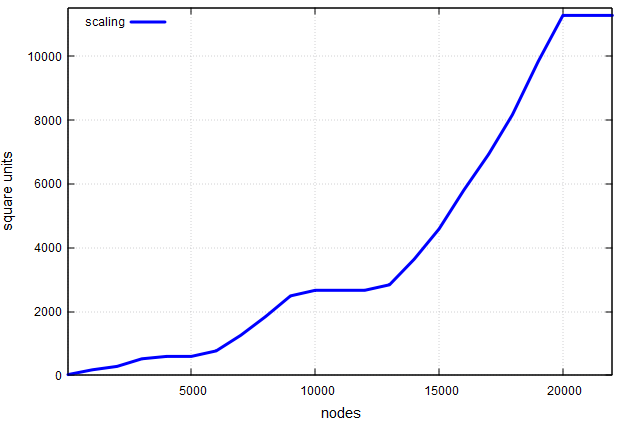}
    \caption{Spatial extent.}
    \label{fig:areasdiamond}
  \end{minipage}
\end{figure}

The heuristic algorithm in 22,030 iterations generates a dense
graph of 123,101 edges with a decaying scaling exponent to 1.1720, Figure~\ref{fig:asymptotic-dia}. Although the graph initially exhibits high density, gradually tends to the theoretical predictions; unlike the greedy version. Figure~\ref{fig:areasdiamond} illustrates the expansion of the search area, which is significantly larger compared to the area of the greedy algorithm and indicates lower density. In total the execution time was more than 24 hours.

\section{Conclusions and future work}
\label{conclusions}

The investigation of the unit distance problem requires the assessment of highly dense and complex graphs. 
As such, working with predefined structures limits our ability to discover the intricate, possibly nearly chaotic, patterns that form unit-distance graphs. Greedy search on the plane enables an unbiased exploration of unit-distance graph generation, as indicated by our experimental results.

The challenges for future work include further assessment of the results of the presented algorithm and the investigation of potential expansion of this method. Analyzing the generated dataset to identify structural patterns and its internal logic will enhance the theoretical investigation of the problem and provide further substantiation of our method. Note that we have used classical statistical methods and advanced AI techniques without much success in analyzing the structural development of the generated graph.

Future research needs to further investigate how the applied constraints affect generality in the following directions:
\begin{itemize}
    \item Evaluate how working with real numbers would benefit the graph generation and by extension how confiding it is to work exclusively in the rational plane.

    \item Identify the optimal set of denominators that will produce the most effective result.

    \item Find the optimal value, or strategy to determine the optimal value, for parameter $h$ and eventually determine how the expansion of the circular search area affects the produced graph.
\end{itemize}

\section*{AI disclosure}

The development of the presented methodology and the substantiation of this work were solely conducted by the author. The coding of the proposed algorithms, their optimization to exploit high-performance hardware (i.e. GPUs) and the execution of the experimental evaluation were AI-assisted tasks.

\bibliography{sn-bibliography}

@article{erdos1946sets,
  title={On sets of distances of n points},
  author={Erd{\"o}s, Paul},
  journal={The American Mathematical Monthly},
  volume={53},
  number={5},
  pages={248--250},
  year={1946},
  publisher={Taylor \& Francis}
}

@misc{openai,
  author       = {OpenAI},
  title        = {Planar Point Sets with Many Unit Distances},
  howpublished = {\url{https://cdn.openai.com/pdf/74c24085-19b0-4534-9c90-465b8e29ad73/unit-distance-proof.pdf}},
  year         = {2026},
  note         = {Accessed: 2026-06-02}
}

@misc{sawin2026explicit,
  author       = {Sawin, Will},
  title        = {An explicit lower bound for the unit distance problem},
  howpublished = {arXiv preprint arXiv:2605.20579},
  year         = {2026},
  eprint       = {2605.20579},
  archivePrefix={arXiv},
  primaryClass ={math.CO},
  month        = {May}
}

@article{spencer1984unit,
  title={Unit distances in the Euclidean plane},
  author={Spencer, Joel and Szemer{\'e}di, Endre and Trotter, William T},
  journal={Graph theory and combinatorics},
  pages={293--303},
  year={1984},
  publisher={Academic Press, New York}
}

@article{engel2025diverse,
  title={Diverse beam search to find densest-known planar unit distance graphs},
  author={Engel, Peter and Hammond-Lee, Owen and Su, Yiheng and Varga, D{\'a}niel and Zs{\'a}mboki, P{\'a}l},
  journal={Experimental Mathematics},
  pages={1--13},
  year={2025},
  publisher={Taylor \& Francis}
}

@misc{repository,
    title={{Unit-Distance Graphs}},
    howpublished="\url{https://rodispantelis.github.io/UnitDistanceGraphs/}"
}

@article{alon2026remarks,
  title={Remarks on the disproof of the unit distance conjecture},
  author={Alon, Noga and Bloom, Thomas F and Gowers, W Timothy and Litt, Daniel and Sawin, Will and Shankar, Arul and Tsimerman, Jacob and Wang, Victor and Wood, Melanie Matchett},
  journal={arXiv preprint arXiv:2605.20695},
  year={2026}
}

@article{emmerich2026optimizing,
  title={Optimizing Explicit Unit-Distance Lower-Bound Certificates},
  author={Emmerich, Michael},
  journal={arXiv preprint arXiv:2606.03419},
  year={2026}
}

\end{document}